# Three prophylactic interventions to counter fake news on social media

Three interventions to counter fake news

Reducing the diffusion of fake news on social media using psychological inoculation, transaction cost economics safeguarding and media literacy interventions.

David A Eccles

School of Computing and Information Systems, The University of Melbourne

Tilman Dingler

School of Computing and Information Systems, The University of Melbourne

Fake news on Social Media undermines democratic institutions and processes. Especially since 2016, researchers from many disciplines have focussed on ways to address the phenomenon. Much of the research focus to date has been on identification and understanding the nature of the phenomenon in and between social networks and of a rather reactive nature. We propose interventions that focus on individual user empowerment, and social media structural change that is prophylactic (pre exposure), rather than therapeutic (post exposure) with the goal of reducing the population exposed to fake news. We investigate interventions that result in greater user elaboration (cognitive effort) before exposure to fake news. We propose three interventions i) psychological inoculation, ii) fostering digital and media literacy and iii) imposition of user transaction costs. Each intervention promises to illicit greater cognitive effort in message evaluation and reduce the likelihood of creating, sharing, liking and consuming 'fake news'.

**ACM Reference Format:**
David A. Eccles, Tilman Dingler. Three prophylactic interventions to counteract fake news on social media. CHI '21 Extended Abstracts in the Proceedings of the Workshop on Technologies to Support Critical Thinking in an Age of Misinformation (CTAM21), May 8–13, 2021, Yokohama, Japan. © 2021 Copyright held by the owner/author(s).

## 1 INTRODUCTION

Fake news on social media undermines democratic processes and institutions (1). It has played a negative role in elections and referendums in a number of countries and threatens many components of what define civil societies (2). Fake news users the elements of news production with none of the processes and oversight (3). Fake news spreads faster, further and lasts longer than factual information on social media (4). Once users are exposed to fake news it is hard to correct the record due to a number of behavioural (5), psychological (6), and cognitive vulnerabilities (7-10). Social media curated news feeds promote in group and out group attitudes and produce filter bubbles and echo chambers (11) where social media users only believe what they trust, rather than trust what is factual. Many researchers are concerned it has created a 'post truth' era will truth and facts no longer play a role (12).

The phenomenon of 'fake news' on social media came to the world's attention in 2016 with two events, first the Brexit Referendum (13) and secondly the 2016 U.S. Presidential election (14). While there were previous disinformation and misinformation campaigns, most notably in Ukraine (15), 2016 was the year that researchers across multiple disciplines focused on the problem. A cross discipline 16 author article published in 'Science' was a call to action to "promote interdisciplinary research to reduce the spread of fake news and to address the underlying pathologies" (3). The call to action proposed two research interventions, the first was individual empowerment of social media users, and the second was structural changes to social media prevent fake news exposure.

The call to action witnessed a cross-discipline response in the social sciences, journalism and the physical sciences (especially computer science, and human computer interaction). Most of this research has been focused at understanding the phenomenon, documenting the fake news process, automated and algorithmic defense and detection, and rediscovery of known human failings based on previous phenomenon such as the MMR (measles, mumps rubella) vaccine anti vaccination movement (16), conspiracy theories, and climate change denial (17). At the same time, social media companies responded by reducing API access to researchers and journalists, avoiding accountability, misdirection and doctoring evidence removing it of context and meaning (18)Much of the efforts to date rely on fact checking which is not without criticism and limitations (19). To date, there has been limited and preliminary efforts in reducing its spread and enabling users to understand fake news construction, enabling user detection and address the population size of affected social media users.

## 2 BACKGROUND

In the following section we outline three approaches to empowering individuals, psychological inoculation, digital media literacy and imposition of safeguarding mechanisms on user exchanges. In our work, we apply two individual interventions, psychological inoculation and digital and media literacy, and one structural change, the imposition of transaction cost safeguarding mechanisms on users (not their posts) involved in exchanging fake news.

### 2.1 2.1 Psychological Inoculation

All humans are subject to social influence through persuasion and propaganda to change their behaviours and attitudes on issues (20). Researchers have been studying how to confer resistance to persuasive messages, and one of the more successful theories is psychological inoculation. Psychological inoculation seeks to 'inoculate' individuals from persuasive messages. It seeks to work in the same way as medical inoculation in conferring resistance to a virus by injecting individuals with an attenuated version of the virus and providing a defence to a future viral attack (21).

Psychological inoculation works by announcing that there will be an attack on an individual's position on an attitude or on their beliefs. This is known as the threat. Once the threat is announced individuals are presented with weakened forms of the message attack. Once the weakened attacks are presented individuals are encouraged to develop refutational preemption to develop defences to the weakened attack (21). Refutation preemption could be active - the individual's initiate the preemption on their own, or passive - individuals use materials provided to them defend their position. Original research used refutation same preemption - meaning the defence against the attack was the same as the weakened attack topic or topics. However later research also extended the psychological inoculation to use refutation different defences, that is defending the position by using arguments different to the content of the weakened attack. Originally there was then a delay period (between hours and weeks) to allow the refutation preemption to ferment before an actual attack on attitudes or beliefs was then presented to the individual (22).

Since 1961 research has extended and varied the psychological inoculation method, and it is generally agreed that the only two elements are essential to psychological inoculation – i) the threat of the attack and ii) the refutational pre-emption (22). The delay element - critical to medical inoculation is not essential to psychological inoculation, and if the delay is greater than fourteen days between the psychological inoculation and the attack the effectiveness of the inoculation is significantly reduced (23).

**2.2  Digital Media Literacy**

Media literacy education, which empowers users with the necessary critical thinking, creative and analytical skill to all forms of media messages, has now seen attempts to extend programs to digital media literacy as well as previous media forms (print, radio, audiovisual(24). Finland (25) and Sweden (26)have long been target of nation state information warfare campaigns from the former Soviet Union and now Russia, including recent moves to digital enviroments. Media and digital literacy skills enable informed engagement in a media rich society. It seeks active engagement with messages rather than non critical passive participation(27).

The ubiquity of ICT devices has put powerful digital tools capable of creating engaging and affecting messages. But not all ICT device users have training on effective tool use. Access to the technology of digital media does not imply that use of that digital media is competent and skilful. As a result media - especially new media forms, such as social media, have influenced large numbers of uncritical unskilled users on their platforms with ongoing threats to democracies across the world~\cite(1). While evidence of self-expression and participation do exist in digital media - critical thinking skills among the majority of users appears to be absent(10). Media literacy seeks to equip citizens to analyse, create, interpret, observe and engage with media messages seeking to influence and persuade (28).

Media Literacy has five key concepts:. First all media messages are constructed. This means that they have structure and form, like a building, a song. Second, that message construction has its own grammar, syntax, tools and materials. Third, different people experience the same message differently. Some people like some songs, some movies, some buildings more than others. Fourth, Media messages represent a specific point of view. The message is telling a story, but in telling a story it is also editing that story and actively omitting other points of view. Fifth - and last, and importantly, media messages are organised to gain advantage. The advantage can be influence, profit or power~\cite(24). As all ICT users are capable of production and consumption, media literacy seeks to apply critical thinking not only to the consumption of media messages, but also to be applied in the creation of media messages and its distribution and production.

**2.3 Transaction Cost Economics**

In any exchange of goods, services or information there are operational costs and transaction costs. While operational costs are the most important in making the decision, the imposition of formal transaction governance often adds unnecessary costs and reduces the effectiveness of the exchange. Transaction Cost Economics (29) is a behavioural economic theory that decides whether the good or service is kept within the firm, or sourced from the market, or partially sourced from the market and the firm (hybrid). It aims to avoid legalistic and legislative governance in the form of an informal governance and dispute mechanism between the parties involved in the transaction.

TCE rests on two assumptions and posits three key dimensions by which the transaction is categorised. The two assumptions are bounded rationality and opportunism, and the three key dimensions are asset specificity, transaction frequency and uncertainty. Bounded rationality is the idea that humans will behave rationally but never 100 percent rationally. The degree of irrational behavior is fluid and will change dependent on the circumstances. The second assumption is that opportunism, 'is a deep condition of self-interest seeking that contemplates guile. Promises to behave responsibly that are unsupported by credible commitments will not, therefore, be reliably discharged' (Williamson, 1998, pp. 68). What Williamson means by opportunism is that if there is no credible and reliable agreement between a vendor and client, then both parties will behave in their own self-interest and this cannot be prevented. Williamson, in this sense is talking about a behavioural contract more than a tort contract to control the transaction (29).

Jones and colleagues (30) expanded Transaction Cost economics hybrid governance to include network governance. This expansion extends Transaction Cost Economics beyond a dyadic exchange to a systems perspective. This system perspective makes Transaction Cost Economics a suitable model for imposing a governance framework on exchanges of information in online social networks. Network governance in exchanges involving multi node rather than dyadic exchanges create exchange problems which can be safeguarded and coordinated by social mechanisms from members in the network (30). While coordination of online social networks lies with the platform providers, safeguarding can be deployed by nodes in the network through social mechanisms. Jones and colleagues identified three social mechanisms to resolve exchange problems: restricted access; collective sanctions and reputation ranking. For restricted access and reputation the amount of restriction and reputation ranking needs to find the sweet spot, too little or too much safeguarding impacts the effectiveness of the network governance.

Media literacy, psychological inoculation and safeguarding assets in transaction cost economic theory are defensive actions designed to protect an asset or attitude from opportunistic behaviour. Each responds to a threat with a proactive defense utilising skills acquisition and literacy, refutation defence or explicit safeguarding. Each method prepares users with the ability to defend themselves from subsequent fake news attacks. In the next section we propose testing these approaches on fake news

**3 METHOD**

In this section we provide an overview of our research focus on user empowerment interventions.

Our first experiment will test social media literacy and psychological inoculation. Psychological inoculation works by giving participants a chance to defend their attitude on an issue by rehearsing a refutation of an attack. Acker and Donovan's (2019) data craft (provides a useful lens to demonstrate psychological inoculation. They identify five elements common to many fake news campaigns: accounts, recycling, content, followers and authentic interactions (18). Digital media literacy focuses on remaining curious and understanding how social media platforms tools enable the construction of information and questioning the purpose, objectives and motives for creating, producing and sharing messages (27).

We can inoculate users by demonstrating characteristics of fake news accounts and content. Accounts often use default avatars,,recent account creation dates, generic biographies, account names with random numbers, and sometimes use AI generated faces for the profile picture (known as GAN images). Fake news accounts recycle account information across several social media platforms (e.g. Twitter, Facebook and Instagram). Fake news accounts post post generic not specific details and context. Images are from other sources and are often not recent and omit recent changes in the urban environment (e.g. building demolitions, road closures). Fake news account followers appear to be clones, often with AI generated profile pictures, generic profiles and non sensical interactions with each other (18). The interactions are designed to algorithmically hijack or promote hashtags and topics (1). By equipping users with the characteristics of fake news accounts and content on social media, users can organise how to approach attacks on the source validity and content on issues by checking account contents and characteristics. Acker and Donovan (2019), provide two examples of this approach in their paper utilising debunked information from the "Black Lives Matter" movement.

Digital medica literacy empowers uses to understand how media is created, produced, shared and consumed. It promotes an active curiosity about all messages and their authorship. Digital media has a grammar and form like all previous media forms (27).

Message creators and producers construct messages, and different social media users will deconstruct the same message in very different and valid ways. Messages represent a point of view and are usually designed to obtain influence, power or financial gain (24). Media literacy asks of users to consider basic questions before liking, sharing or commenting on information they see in their social media feeds. Media literate users interrogate the message form as well as its content with a series of questions. Some examples of questions users may not have previously consider include: Are the sources reliable and with relevant knowledge on the topic? How would you fact check this given your search engine results are customised to your wants and may not present salient and relevant information? Is the information current or old? Is the message in context or is it a snippet taken out of context? It also asks uses to consider the motives of the message creator and account sharing the information. What is there motive in sharing? What points of view are missing from this message? Critically given fake news is designed to illicit negative affected states why are you affected by this message? Why does the message appeal to you or upset you? What is your perspective on this issue, and what would be a valid different attitude to your own? It is important to stress that this is not about being cynical about messages, but about objective inquiry and curiosity (24 , 28). Digital media literacy seeks to promote deeper critical thinking about digital tools, message authorship, construction, and consumption.

Our second experiment will ask participants to test a structural change to social media by the imposition of safeguarding mechanism on the transaction - part of transaction cost economic theory (31). Safeguarding is a mechanism used to protect assets. Assets of low specificity require no safeguarding, however once the asset has unique attributes and characteristics, the failure to safeguard is considered an unrelieved hazard (29). In network environments like social media safeguarding takes the form of reputation ranking, temporary user suspension, and restricting user access to other users. To date, social media fake news interventions focus on the content (post), not the node (user). Transaction cost asset safeguarding focuses on the nodes in a social network, not the information shared across the dyadic tie between nodes. By imposing a reputation ranking on a node in the network, other nodes may restrict information and resources they will share with the node and share from that node with other users.

## 4 DISCUSSION

All of the research efforts to date give researchers and the broader community greater understanding of the problem, and the magnitude and variety of the significant threats to individuals, social groups, organisations, governments and societies. The user's experience of responses to fake news on social media is crowdsourced, and media organisation fact checking of posts. By providing fact checking and editorial mechanisms these organisations are restoring news media processes and oversight missing from fake news. However, the focus on the content of fake news, does not empower social media users nor encourages structural changes. Critically, the majority of responses to the fake news pathology do not prevent user exposure to fake news, nor ameliorate the rapid diffusion to a vast social media user population. The research that does investigate preventative approaches to fake news self identify their work as being in the preliminary stages. Most research is post exposure to the fake news pathology, leaving social media users vulnerable.

The significant majority of research focuses on the message, but not the message producer, creator and disseminators. Contagion in social networks spreads from node to node, and the ties that connect the nodes is the conduit for fake news dissemination. Empowering the nodes of a social network, to detect fake news, and evaluate the reputation of the node source sharing fake news may change the receiving nodes attitude to further distribution of fake news within the network.

Social media users have two states when it comes to fake news. User can exist in a pre exposure state, or a post exposure state. Empowering users post exposure to fake news often means the damage is done, the fake news has done its job and successfully exploited the various human vulnerabilities in message processing. Users may be exposed to messages that utilise psychological profiling, or manipulate them to be in a negative affected state. Correcting the record with factual information may not be sufficient to reverse the effects of fake news. Further challenges arise with humans trusting the first message on a topic more than subsequent messages (32), so if fake news is not countered in a sufficient time frame or multiple attitudes on the topic are not visible - and result in greater cognitive effort - heuristics will form based on the first message and lazy cognitive processing will be employed in subsequent messages which are evaluated against the first message processed.

We intend to use psychological inoculation, digital media literacy and imposition on transaction costs on users (not their messages) as a prophylactic (pre-exposure) treatment to empower social media users. Each of these treatments empowers users and does not require users to encounter fake news unprepared. Knowing why the message has put a user in an affected state  understanding the construction of the message, and that the message author has a poor reputation for fake news are three examples of the kind of empowered social media user we envisage. This has many benefits, it is platform independent, scalable, and applicable to social media users of all cognitive abilities and ages.

There are several limitations to this paper, namely, social media access, system visibility and environmental shaping. Social media companies are restricting and curating access, making it harder for researchers to study the phenomenon. Younger users are moving to platforms that have no API access or moving from public broadcast to private narrowcast (e.g. WhatsApp private message groups) removing visibility. Interventions, particularly social media literacy and psychological inoculation may suffer the same problem as nudging. By empowering individuals, we risk training them to create better fake news campaigns. The last is the distinction between virtual and physical worlds. Cognitive elaboration in evaluation of persuasive messages and propaganda is successful in face to face encounters, however, there is limited peer reviewed evidence supporting greater collaborative elaboration works in online social networks like social media. Future research should consider the efficacy and longevity of pre and post exposure interventions and the most effective way to scale and automate successful interventions.

## 5  CONCLUSION

Fake news on social media targets individuals, attacks social groups and undermines democratic processes and democratic institutions. In this paper we consider the research approaches to date as well as propose three prophylactic interventions on individual social media users: i) psychological inoculation ii) imposition of transaction costs between user exchanges and iii) digital media literacy. We hypothesize that these interventions will result in greater cognitive effort in message elaboration resulting in users identifying fake news messages and as a result choosing not to share or like fake news. The reduction in promoting fake news could theoretically reduce the organic and algorithmic spread within online social networks.